\documentclass{article}
\usepackage{ijcai25}

\usepackage{times}
\usepackage{soul}
\usepackage{url}
\usepackage[hidelinks]{hyperref}
\usepackage[utf8]{inputenc}
\usepackage[small]{caption}
\usepackage{graphicx}
\usepackage{amsmath}
\usepackage{amsthm}
\usepackage{booktabs}
\usepackage{algorithm}
\usepackage{algorithmic}
\usepackage[switch]{lineno}
\usepackage{amsfonts}
\usepackage{multicol}
\usepackage{multirow}
\usepackage{threeparttable}
\usepackage{float}
\usepackage{makecell}

\title{Enhance Generation Quality of Flow Matching V2A Model via Multi-Step CoT-Like Guidance and Combined Preference Optimization}

\author{
Haomin Zhang$^1$\and
Sizhe Shan$^2$\and
Haoyu Wang$^2$\and
Zihao Chen$^1$\and
Xiulong Liu$^3$\and
Chaofan Ding$^1$\and
Xinhan Di$^1$\\
\affiliations
$^1$AI Lab, Giant Network.\\
$^2$Zhejiang University\\
$^3$University of Washington\\
\emails
\{zhanghaomin, chenzihao, dingchaofan, dixinhan\}@ztgame.com,\\
\{22360174, 22331169\}@zju.edu.cn,
xl1995@uw.edu
}

\begin{document}

\maketitle

\begin{abstract}
Creating high-quality sound effects from videos and text prompts requires precise alignment between visual and audio domains, both semantically and temporally, along with step-by-step guidance for professional audio generation. However, current state-of-the-art video-guided audio generation models often fall short of producing high-quality audio for both general and specialized use cases. To address this challenge, we introduce a multi-stage, multi-modal, end-to-end generative framework with Chain-of-Thought-like (CoT-like) guidance learning, termed Chain-of-Perform (CoP). First, we employ a transformer-based network architecture designed to achieve CoP guidance, enabling the generation of both general and professional audio. Second, we implement a multi-stage training framework that follows step-by-step guidance to ensure the generation of high-quality sound effects. Third, we develop a CoP multi-modal dataset, guided by video, to support step-by-step sound effects generation. Evaluation results highlight the advantages of the proposed multi-stage CoP generative framework compared to the state-of-the-art models on a variety of datasets, with FAD $0.79$ to $0.74$ ($+6.33\%$), CLIP $16.12$ to $17.70$ ($+9.80\%$) on VGGSound, SI-SDR $1.98 dB$ to $3.35 dB$ ($+69.19\%$), MOS $2.94$ to $3.49$ ($+18.71\%$) on PianoYT-2h, and SI-SDR $2.22 dB$ to $3.21 dB$ ($+44.59\%$), MOS $3.07$ to $3.42$ ($+11.40\%$) on Piano-10h.

\end{abstract}

\section{Introduction}
Foley, the art of synthesizing ambient sounds and sound effects guided by videos, aims to produce high-quality audio, such as background music or human speech, that meets two essential requirements: (1) semantic alignment and (2) temporal synchronization with the associated videos. Foley methods are expected to understand scene contexts and their relationship with audio, while also ensuring audio-visual synchronization, as humans are highly sensitive to mismatches between sound and visuals ~\cite{luo2023difffoley,zhang2024foleycrafter,SpecVQGAN_Iashin_2021,wang2024frieren,cheng2024taming}. Existing Foley models can be categorized into two main groups. The first group concentrates on improving alignment using specially designed modules ~\cite{zhang2024foleycrafter,wang2023v2amapperlightweightsolutionvisiontoaudio,li2024tri}, The second group seeks to enhance alignment performance by utilizing a unified DiT model architecture~\cite{cheng2024taming,chen2024video}. However, existing video-to-audio (V2A) models often face challenges in achieving accurate alignment across both semantic and temporal domains when guided by visual information. To overcome these limitations, we propose a data-driven approach for synthesizing high-quality audio that enhances both semantic and temporal alignment beyond the scope of traditional supervised learning.

Besides, current state-of-the-art V2A methods neither train on audio-visual data of professional audio aspects such as piano, violin, and movie effects~\cite{Lee2019ObservingPA,Koepke2020SightTS} nor follow professional step-by-step guidance in the learning process to produce high-quality professional audio ~\cite{Su2020AudeoAG}. Furthermore, it is unclear whether pre-trained V2A models designed for generating common audio can effectively support video-to-professional audio scenarios, such as generating audio for a piano performance.

Therefore, to further improve the alignment between visual and audio in the V2A (common audio) synthesis that jointly considers video, audio, and text in a transformer-based network. We apply a post-training contrastive learning strategy in order to factor multiple domains into shared and unique representations~\cite{liang2024factorized}. Then, inspired by the Chain-of-Thought~\cite{wei2022chain} in large language models (LLMs), we propose CoP to support the step-by-step generation of professional audio (piano) on the basis of the pre-trained V2A (common) model. Finally, to capture various playing styles in professional audio, we apply direct preference optimization (DPO)~\cite{rafailov2024directpreferenceoptimizationlanguage}, extending beyond dense-labeled supervised learning. 

In summary, we propose a multi-modal, multi-stage paradigm via CoP for V2A (common and professional audio) generation. In the three-stage training stages, both dense-labeled supervised learning and combined preference learning are conducted. Besides, we propose CoP guidance to support high-quality professional audio generation from corresponding visual inputs (videos). Finally, a step-by-step CoP multi-modal dataset is built for the Video-to-Piano audio generation.

\section{Related Work}

\subsection{Video-to-Audio}
With the rapid development of video generation and Text-to-Audio (T2A) technologies, the task of adding Foley effects to silent videos has attracted increasing attention. Some studies~\cite{SpecVQGAN_Iashin_2021,viertola2024temporally,mei2024foleygen} have adopted autoregressive methods to generate audio tokens, which are then decoded into audio signals. Meanwhile, the robust capabilities of latent diffusion and flow matching techniques have substantially enhanced both the quality and efficiency of foley production for silent videos~\cite{luo2023difffoley,wang2024frieren}. Some works~\cite{zhang2024foleycrafter,li2024tri} have introduced additional control conditions, such as timestamp and energy, to improve audio quality. MultiFoley~\cite{chen2024video} combines mask denoising with reference audio to achieve video-guided, multi-modal control over the audio generation and extension. Moreover, MMaudio~\cite{cheng2024taming} utilizes a multi-modal transformer to perform audio generation through flow matching, incorporating a synchronization module to effectively enhance audio-video temporal alignment. However, existing studies have not adequately addressed the modality differences between audio and video, nor have they thoroughly explored CoT-like guidance in this context. Therefore, building on the DiT flow matching model~\cite{liu2022flow}, we utilize FactorCL to enhance the alignment between different domains and propose a CoT-like V2A method. This approach facilitates the generation of both general audio (e.g., VGGSound~\cite{chen2020vggsound}) and professional audio (e.g., piano performances) by leveraging step-by-step guidance through a CoP mechanism.

\subsection{Visual Piano Transcription}
Research in video-based piano Automatic Music Transcription (AMT) has evolved significantly over the years, primarily focusing on predicting MIDI from video. Recent work~\cite{Lee2019ObservingPA,Koepke2020SightTS,Su2020AudeoAG} has seen widespread adoption of CNN approaches to predict pitch onset events from video frame sequences. Audeo~\cite{Su2020AudeoAG} introduced a three-stage pipeline for generating audio from silent piano performance videos. Their approach utilized an enhanced ResNet~\cite{he2015deepresiduallearningimage} to predict pitch onset-offset events from video frames, then refined the MIDI predictions through a GAN network~\cite{2014Generative}, and finally converted the predicted MIDI to audio using a MIDI synthesizer. These previous methods generate piano sounds through MIDI predictions, however, current methods are focused on a narrow domain. Furthermore, it remains unclear whether pre-trained V2A models designed for generating common audio can benefit professional audio generation scenarios. Therefore, we propose CoP guidance and a corresponding multi-modal (Visual-Audio-Text-MIDI) network architecture, built upon a flow matching based V2A model.
      

\begin{figure*}[t]
    \centering    \includegraphics[width=0.75\textwidth]{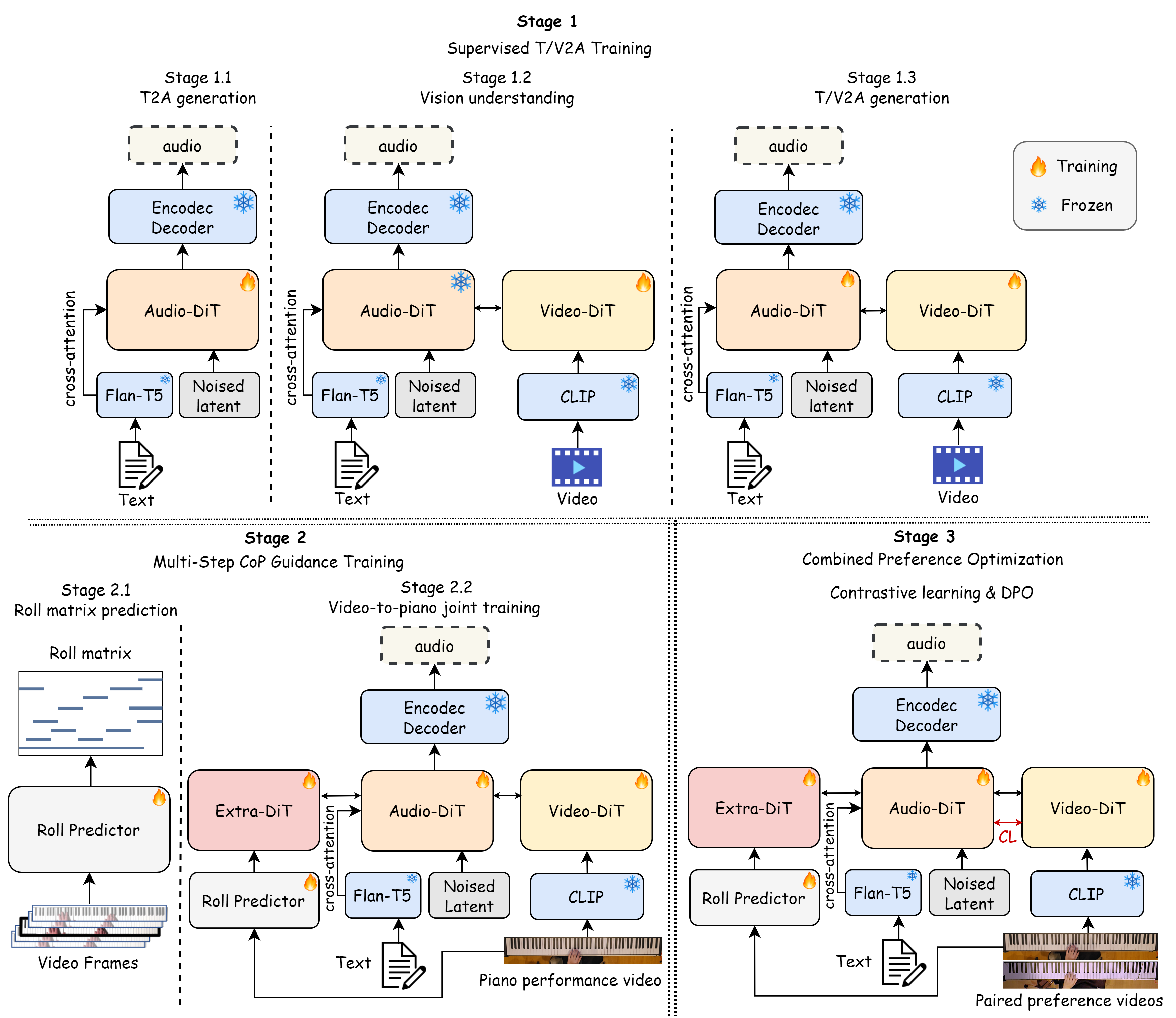}
    \caption{Multi-stage training pipeline of our method.}
    \label{fig:architecture}
\end{figure*}

\subsection{Multi-modal Chain-of-Thought Reasoning and Generation}
Multi-modal reasoning and generation tasks require models to possess multi-domain perception and high-level cognitive abilities ~\cite{xu2024llava}. Multi-modal large language models (MLLMs) are excepted to answer complex questions requiring reasoning ~\cite{zhang2024improve}. CoT is explored to enhance the ability of reasoning in MLLMs for tasks of understanding and generation~\cite{xu2024llava}. A variety of CoT-like methods are studied to enhance the ability of reasoning for the MLLMs ~\cite{rafailov2024direct,xu2024llava}. However, step-by-step guidance is not studied in high-quality audio generation from visual inputs. Therefore, we propose a CoT-Like (CoP) guidance learning and corresponding network architecture to improve the generation quality of both general (VGGSound ~\cite{chen2020vggsound}) and professional audio (piano) based on flow matching transformers.      


\section{Method}
\subsection{Overview}
A multi-modal multi-stage paradigm with CoP guidance for V2A (common and professional audio) generation is represented here. A corresponding three-stage training process is conducted. In the first stage, a flow matching ~\cite{liu2022flow} based transformer to perform V2A and T2A (common audio from VGGSound~\cite{chen2020vggsound} and T2A Datasets) tasks is trained. In the second stage, multi-step CoP guidance training is conducted to improve the generation quality of professional audio (piano). In the third stage, combined preference learning is applied to enhance the alignment between visual and audio representations while improving the generation quality of professional audio (Piano).
The three training stages are represented as the following:
\begin{itemize}
    \item \textbf{Stage 1:} Supervised T/V2A Training.
    
    \item \textbf{Stage 2:} Multi-Step CoP Guidance Training.
    
    \item \textbf{Stage 3:} Combined Preference Optimization.
\end{itemize}

%

The detailed formulation is provided as follows:
\begin{small}
\begin{equation}\label{eq_overview}
\begin{aligned}
    F^{\text{final}} &= \mathbb{F}^{\text{multi\mbox{-}stage}}(C_{\text{text}}, C_{\text{video}}, G_{\text{det}}, G_{\text{cl}}, G_{\text{pre}}) \\
    &= \mathbb{F}^{\text{stage3}}(\mathbb{F}^{\text{stage2}}(\mathbb{F}^{\text{stage1}}(C_{\text{text}}, C_{\text{video}}), G_{\text{det}}), G_{\text{cl}}, G_{\text{pre}})
\end{aligned}
\end{equation}
\end{small}
where $C_{\text{text}}, C_{\text{video}}$ are the input text and video conditions, $G_{\text{det}}, G_{\text{cl}}, G_{\text{pre}}$ are the guidance of detailed information (like music melody), contrastive learning and preference data respectively, $\mathbb{F}(.)$ denotes the training process of different stages, and $F^{\text{final}}$ is the final model we get.


\subsection{Stage 1: Supervised T/V2A Training}
\begin{small}
\begin{equation}\label{eq_stage1}
\begin{aligned}
    l_{\text{v}}, & e^{\text{stage1}}_{\text{Audio\mbox{-}DiT}} = F^{\text{stage1}}_{\text{Audio\mbox{-}DiT}}(l_{\text{n}}, F_{\text{Flan\mbox{-}T5}}(C_{\text{text}}),e^{\text{stage1}}_{\text{Video\mbox{-}DiT}}) \\
    & e^{\text{stage1}}_{\text{Video\mbox{-}DiT}} = F^{\text{stage1}}_{\text{Video\mbox{-}DiT}}(F_{\text{CLIP}}(C_{\text{video}}),e^{\text{stage1}}_{\text{Audio\mbox{-}DiT}}))
\end{aligned}
\end{equation}
\end{small}

In stage 1, we perform audio generation leveraging rectified flow matching~\cite{liu2022flow} based on a multi-stream DiTs~\cite{peebles2023scalable} architecture as described like Eq.~\ref{eq_stage1}, where \(l_\text{v}\), \(l_\text{n}\) are Encodec~\cite{defossez2022high} audio latent for flow matching velocity and noisy input respectively, \(e\) denotes the output of the corresponding DiT in every layer, and \(F\) denotes the corresponding module. Different modalities and DiTs are trained step-by-step as shown in Fig.~\ref{fig:architecture}.

Flow matching calculates the probabilistic paths by predicting the noise distribution to the probability vector field of audio latent. The loss function is defined as follows:
\begin{equation}\label{flowmatching}
\mathcal{L}_\textit{CFM}(\theta) = \mathbb{E}_{t,p_1(x_1), p_t(x|x_1)} ||v_{\theta}(x,t) - u_t(x|x_1)||^2
\end{equation}
We perform the flow matching process in the latent space, using Encodec~\cite{defossez2022high} to obtain audio latent, which are pre-trained on 24 kHz monophonic audio across various domains, including speech, music, and general audio. Specifically, we utilize features extracted before the residual quantization layer.

Our model supports conditioning on both text prompts and videos. To enable effective cross-modality mapping, we employ an audio-video mapping module to fuse the outputs of Audio-DiT and Video-DiT at each layer. Outputs of the DiTs, \(e_a\), \(e_v\) are concatenated and linearly projected before being added to the original outputs, following \(e_a^{'} = e_a + \text{Linear}_a(\text{concat}(e_a, e_v))\). For the text-conditioned audio generation, following the method of Tango~\cite{ghosal2023tango}, we employ an instruction-tuned large language model, FLAN-T5~\cite{chung2024scaling}, as the text encoder and incorporate cross-attention in each layer of the Audio-DiT. For video-conditioned audio generation, we use CLIP~\cite{radford2021learning} as the visual encoder to extract frame-level features, which are then upsampled to match the length of the audio frames.

\subsection{Stage 2: Towards Multi-Stage Chain-of-Perform Guidance Training}

\begin{small}
\begin{equation}\label{eq_stage2}
\begin{aligned}
    l_{\text{v}}, & e^{\text{stage2}}_{\text{Audio\mbox{-}DiT}} = F^{\text{stage2}}_{\text{Audio\mbox{-}DiT}}( l_{\text{n}}, F_{\text{Flan\mbox{-}T5}}(C_{\text{text}}),e^{\text{stage2}}_{\text{Video\mbox{-}DiT}},e^{\text{stage2}}_{\text{Extra\mbox{-}DiT}}) \\
    & e^{\text{stage2}}_{\text{Video\mbox{-}DiT}} = F^{\text{stage2}}_{\text{Video\mbox{-}DiT}}(F_{\text{CLIP}}(C_{\text{video}}),e^{\text{stage2}}_{\text{Audio\mbox{-}DiT}}), \\
    & e^{\text{stage2}}_{\text{Extra\mbox{-}DiT}} = F^{\text{stage2}}_{\text{Extra\mbox{-}DiT}}(F^{\text{stage2}}_{\text{Roll\mbox{-}Predictor}}(C_{\text{video}}),e^{\text{stage2}}_{\text{Audio\mbox{-}DiT}})) \\
\end{aligned}
\end{equation}
\end{small}

In Stage 2, we introduce two additional modules for the complex, domain-specific tasks like piano generation: Extra-DiT and Roll Predictor as Eq.~\ref{eq_stage2}. Modules are trained step-by-step as shown in Fig.~\ref{fig:architecture}.

The Extra-DiT module is designed to process more precise and detailed information, specifically, the piano roll matrix, which encodes the pitch and duration of piano notes. Similar to the audio-video mapping module, the output of each layer in Extra-DiT is also mapped to Audio-DiT to ensure accurate audio-video alignment.


We employ an improved ResNet model~\cite{Su2020AudeoAG} to predict the piano roll from video frames as the Roll Predictor. To efficiently predict pitch and onset-offset events, we design a piano roll matrix \(M\) as a control signal. Specifically, \(M\) is a two-dimensional binary matrix \(M \in \mathbb{R}^{T \times N}\), where \(T\) is the number of video frames and \(N\) is the number of notes (typically 88) in a piano. In this matrix, the pressed notes in each frame are set to 1, while all other positions are set to 0. In experiments incorporating velocity (note strike intensity) guidance, the values of 1 are replaced with the actual relative velocity values. The model processes five consecutive video frames at a time and predicts all the notes pressed during the middle frame. Roll Predictor is trained using mean squared error (MSE). The predicted roll matrix is then projected to the Extra-DiT as an additional condition. Subsequently, the three DiTs and Roll Predictor are jointly trained on both piano performance data and the original T2A and V2A data. The joint loss function is \( L_{\textit{stage2.2}} = L_{\text{roll}} + L_{\text{fm}} \).

\subsection{Stage 3: Combined Preference Optimization}~\label{factorcl} 

In stage 3, we employ Contrastive Learning and DPO to further enhance sound quality. Contrastive Learning and DPO are applied to models in stage 1 and stage 2, respectively.

\paragraph{Conditional Factorized Contrastive Learning.}
We utilize two different approaches for contrastive learning. The first is Supervised Contrastive Learning~\cite{khosla2020supervised}. We select the outputs from the first layer of the Audio-DiT and Video-DiT, denoted as \(e^a\) and \(e^v\), respectively, as contrastive learning samples. Within a training batch, we randomly sample a series of consecutive frames \(j\) from the clip \(i\), denoted as \(e^a_{ij}\) and \(e^v_{ij}\), where frames from the same clip are considered positive samples, and frames from different clips are considered negative samples. The loss function is defined as follows:

\begin{small}
\begin{equation}\label{eq4}
\begin{aligned}
L_{\text{CL}} = -\sum_{i,j} \log \frac{\sum_{l} \exp{(f_a(e^a_{ij}) \cdot f_v(e^v_{i,l}) / \tau)}}{\sum_{k,l} \exp{(f_a(e^a_{ij}) \cdot f_v(e^v_{k,l}) / \tau)}}
\end{aligned}
\end{equation}
\end{small}

Where \(f_a(\cdot)\) and \(f_v(\cdot)\) are the projection modules for audio and video embeddings respectively, which consist of a linear layer followed by L2 normalization, and \( \tau \) is a scalar temperature parameter.

The second approach uses Factorized Contrastive Learning (FactorCL)~\cite{liang2024factorized}, which is designed for multi-modal scenarios to balance shared and modality-specific information. Unlike traditional contrastive learning, which maximizes only cross-modal mutual information, FactorCL decomposes the total mutual information $I(X_1, X_2; Y)$ between multi-modal data and the task from an information-theoretic perspective into three components: the cross-modal shared information $I(X_1; X_2; Y)$ and the unique information of each modality, $I(X_1; Y\mid X_2)$ and $I(X_2; Y\mid X_1)$. $X_1$ and $X_2$ represent two distinct modalities of input data (e.g., video and audio), and $Y$ denotes the corresponding task label (e.g. label in classification tasks).

FactorCL simultaneously learns four factorized representations $\{Z_{S_1}, Z_{S_2}, Z_{U_1}, Z_{U_2}\}$ to capture ``task-relevant shared information'' (denoted \(S\)) and ``task-relevant unique information'' (denoted \(U\)). This is achieved by estimating mutual information using both lower bounds (InfoNCE) and upper bounds (NCE-CLUB), alongside data augmentation (including unique augmentation) to remove irrelevant noise in an unsupervised setting. The training objective is to maximize the sum of the shared information term \(S\) and the unique information terms \(U_1\) and \(U_2\). The final factorized contrastive loss function \(L_{\textit{FactorCL}}\) is written as Eq.~\ref{eq3}.

\begin{small}
\begin{equation}\label{eq3}
    L_{\textit{FactorCL}} = - (S + U_1 + U_2)
\end{equation}
\end{small}

This method typically outperforms traditional contrastive learning on multi-view non-redundant datasets, achieving higher accuracy and stronger representation in downstream tasks. Based on Eq.\ref{eq3}, we apply FactorCL to \(e_a\) and \(e_v\). \(e_a\) represents noisy audio embeddings, while \(e_v\) corresponds to video embeddings in multi-modal sound generation task. The objective of FactorCL is to improve the generation task by extracting accurate audio information from noisy audio and video embeddings.

\(L_{\text{CL}}\) and \(L_{\textit{FactorCL}}\) are used to post train modules \(F^{\text{stage1}}_{\text{Audio\mbox{-}DiT}}\) and \(F^{\text{stage1}}_{\text{Video\mbox{-}DiT}}\) from stage 1.

\paragraph{Direct Preference Optimization.}

As for the piano music generation task, we further incorporate preference learning to enhance sound quality, particularly for generating the desired playing style. We fine-tune our piano model using preference data from two pianists in the Piano-10h dataset with DPO optimization. We first label the training data, which includes two distinct styles of piano music (e.g. ``preferred sample'' and ``control sample''). The objective is to generate audio that closely resembles the preferred samples while diverging from the control samples, guided by the following loss function:

\begin{small}
\begin{equation}\label{preference}
\begin{aligned}
& L_{\text{dpo}}(\pi_\theta,\pi_{ref}) = \\
& -E_{(x,y_w,y_l)}[log\sigma(\beta\log\frac{\pi_\theta(y_w|x)}{\pi_{ref}(y_w|x)} - \beta\log\frac{\pi_\theta(y_l|x)}{\pi_{ref}(y_l|x)})] \\
\end{aligned}
\end{equation}
\end{small}

where \(y_w,w_l\) are the preferred and control samples respectively, \(x\) is the corresponding condition like MIDI info, \(\pi_{ref}\) is the reference piano model we use.

\(L_{\text{dpo}}\) is used to post train all the modules from stage 2.

The integration of contrastive learning and preference learning enables the base flow matching network to address diverse audio style generation requirements, enhancing sound quality and fine-grained control through our post-optimization.

%
%


\begin{figure*}[!htb]
    \centering
    \includegraphics[width=0.8\textwidth]{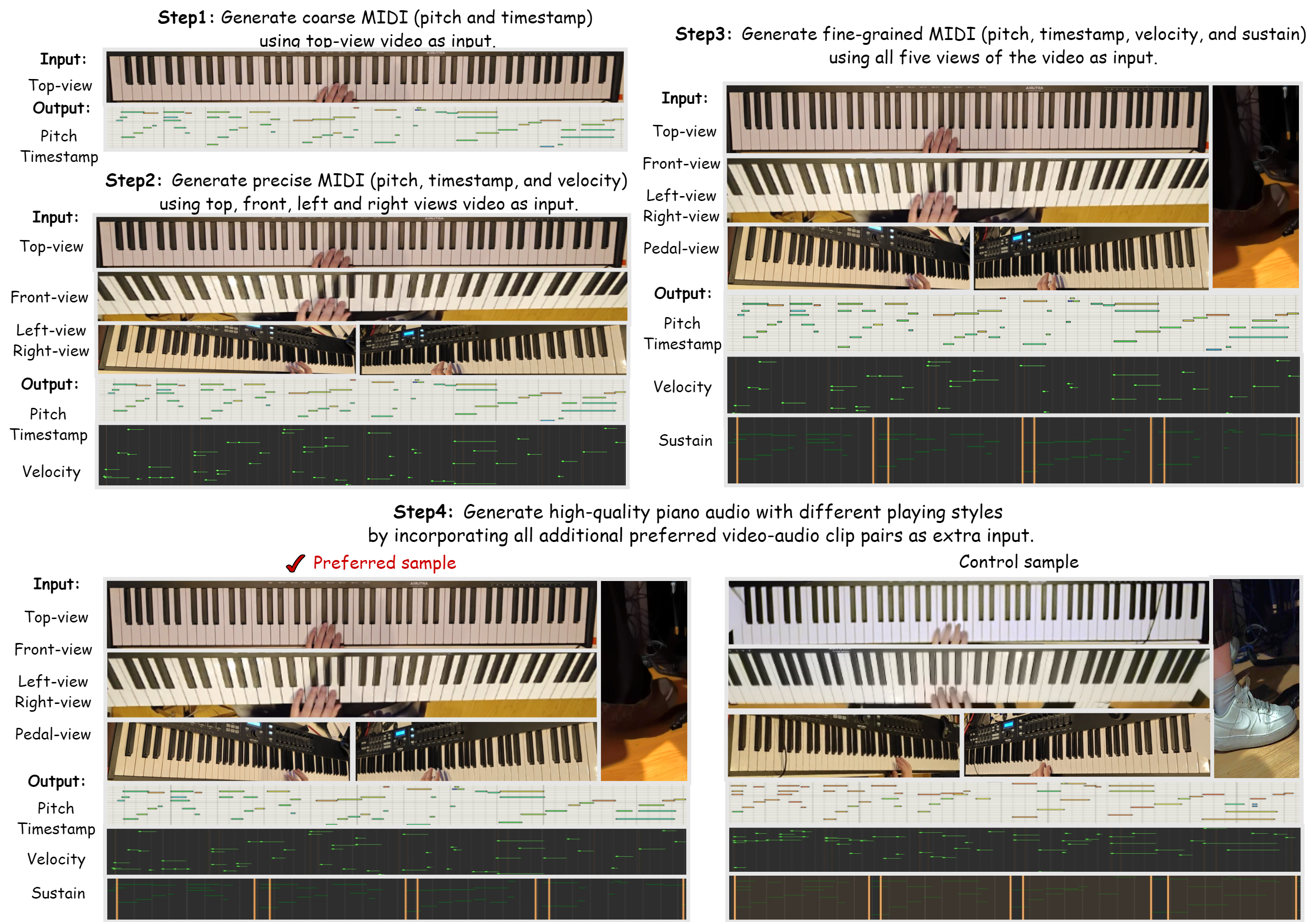}
    \caption{Five views of the Piano-10h dataset supporting step-by-step generation tasks.}
    \label{views}
\end{figure*}

\section{Piano-10h Chain-of-Perform Dataset}

We have constructed a 10-hour multi-modal video-to-piano CoT-like (CoP) dataset for generating high-quality professional piano audio from videos. The primary constraint for data collection was a five-view piano performance with a fully visible keyboard and practice pedal. We employed two skilled pianists with different performance styles to record this dataset. Additionally, a step-by-step CoP guidance was developed. As shown in Fig.~\ref{views}, our expert pianist manually provides step-by-step guidance annotations to support the high-quality generation of piano audio, inspired by CoT and CoT-like guidance \cite{wei2022chain}. Specifically, to generate high-quality piano audio from videos, step-by-step instructions with corresponding ground truth annotations and visual input are provided. As depicted in Fig.~\ref{views}, the process consists of four steps:

\begin{enumerate}
    \item Generate coarse MIDI (pitch and timestamp) using the corresponding top-view video as input.
    \item Generate precise MIDI (pitch, timestamp, and velocity) using the corresponding top, left, right, and front views video as input.
    \item Generate fine-grained MIDI (pitch, timestamp, velocity, and sustain) using all five views of the video as input.
    \item Generate high-quality piano audio with different playing styles by incorporating all additional preferred video-audio clip pairs as extra input.
\end{enumerate}

\begin{table}[h]
\centering
\resizebox{0.9\columnwidth}{!}{%
\setlength{\tabcolsep}{1.0mm}{
\begin{tabular}{lcc}
\toprule
Dataset & Modality & Clips \\
\midrule
AudioCaps~\cite{kim2019audiocaps} & T/A & 49k \\
WavCaps~\cite{mei2024wavcaps} & T/A & 402k \\
TangoPromptBank~\cite{ghosal2023tango} & T/A & 37k \\
MusicCaps~\cite{agostinelli2023musiclm} & T/A & 5k \\
AF-AudioSet~\cite{kong2024improving} & T/A & 695k \\
VGGSound~\cite{chen2020vggsound} & T/V/A & 173k \\
\bottomrule
\end{tabular}}}
\caption{Dataset details.}
\label{tab:datasets}
\end{table}
\begin{table*}[t]
\centering
\resizebox{0.75\textwidth}{!}{%
\begin{tabular}{lccccccc}
\toprule
Method & Params & FAD$\downarrow$ & FD$\downarrow$ & KL$\downarrow$ & IS$\uparrow$ & CLIP$\uparrow$ & AV$\uparrow$\\
\midrule
Diff-Foley~\cite{luo2023difffoley} * & 859M & 6.05 & 23.38 & 3.18  & 10.95 & 9.40  & 0.21\\
FoleyCrafter w/o text~\cite{zhang2024foleycrafter} * & 1.22B & 2.38 & 26.70 & 2.53 & 9.66 & 15.57 & {\bfseries 0.25}\\
FoleyCrafter w. text~\cite{zhang2024foleycrafter} * & 1.22B & 2.59 & 20.88 & 2.28 & 13.60 & 14.80 & 0.24\\
V2A-Mapper~\cite{wang2024v2a} * & 229M & 0.82 & 13.47 & 2.67 & 10.53 & 15.33 & 0.14 \\
Frieren~\cite{wang2024frieren} * & 159M & 1.36 & 12.48 & 2.75 & 12.34 & 11.57 & 0.21 \\
MMAudio-S-16kHz~\cite{cheng2024taming} & 157M & 0.79 & 5.22 & {\bfseries 1.65} & 14.44 & - & - \\
MMAudio-S-44.1kHz~\cite{cheng2024taming} & 157M & 1.66 & 5.55 & 1.67 & {\bfseries 18.02} & - & - \\
MMAudio-M-44.1kHz~\cite{cheng2024taming} & 621M & 1.13 & 4.74 & 1.66 & 17.41 & - & - \\
MMAudio-L-44.1kHz~\cite{cheng2024taming} & 1.03B & 0.97 & {\bfseries 4.72} & {\bfseries 1.65} & 17.40 & 16.12 * & 0.22 * \\
\midrule
Ours-Base w/o text & 711M & 0.80  &  8.66    & 2.22 & 12.08 & 16.14 & {\bfseries 0.25}\\
Ours-Base w. text & 711M & 0.78 & 6.28 & 1.73 & 14.02 & 16.86 & {\bfseries 0.25}\\
Ours-Piano2h w. text & 789M & 0.83 & 6.97 & 1.74 & 13.99 & 16.49 & 0.23 \\
Ours-CL w. text & 712M & 0.75 & 6.42 & 1.70 & 14.72 & 17.09 & 0.24 \\
Ours-FactorCL w. text & 718M & {\bfseries 0.74} & 5.69 & 1.69 & 14.63 & {\bfseries 17.70} & 0.24 \\
\bottomrule
\end{tabular}}
\caption{Objective results of VGGSound-Test regarding audio quality, semantic and temporal alignment. w. text denotes audio generation with text as a guiding condition, and w/o text denotes audio generation without text input, using only the video input. *: These are reproduced using their official checkpoints and inference codes, following the same evaluation protocol.}
\label{tab:vggsound}
\end{table*}

\begin{table*}[htb]
\centering
\resizebox{0.6\textwidth}{!}{%
\begin{tabular}{lccccc}
\toprule
Method & Params & FAD$\downarrow$ & FD$\downarrow$ & IS$\uparrow$ & CLAP$\uparrow$ \\
\midrule
AudioLDM2-L~\cite{audioldm2-2024taslp} & 712M & 5.11 & 32.50 & 8.54 & 0.212 \\
TANGO~\cite{ghosal2023tango} & 866M & 1.87 & 26.13 & 8.23 & 0.185 \\
TANGO2~\cite{majumder2024tango} & 866M & 2.74 & 19.77 & 8.45 & 0.264 \\
Make-An-Audio~\cite{huang2023make} & 453M & 2.59 & 27.93 & 7.44 & 0.207 \\
Make-An-Audio2~\cite{huang2023make2} & 937M & 1.27 & 15.34 & 9.58 & 0.251 \\
GenAU-Large~\cite{haji2024taming} & 1.25B & {\bfseries 1.21} & 16.51 & 11.75 & 0.285 \\
MMAudio-S-16kHz~\cite{cheng2024taming} & 157M & 2.98 & 14.42 & 11.36 & 0.282 \\
MMAudio-S-44.1kHz~\cite{cheng2024taming} & 157M & 2.74 & 15.26 & 11.32 & 0.331 \\
MMAudio-M-44.1kHz~\cite{cheng2024taming} & 621M & 4.07 & {\bfseries 14.38} & 12.02 & {\bfseries 0.351} \\
MMAudio-L-44.1kHz~\cite{cheng2024taming} & 1.03B & 4.03 & 15.04 & {\bfseries 12.08} & 0.348 \\
\midrule
Ours-Base & 711M & 1.48 & 18.10 & 9.65 & 0.334 \\
\bottomrule
\end{tabular}}
\caption{Objective results of AudioCaps-Test.}
\label{tab:audiocaps}
\end{table*}

\begin{figure}[!htb]
    \centering    \includegraphics[width=0.35\textwidth]{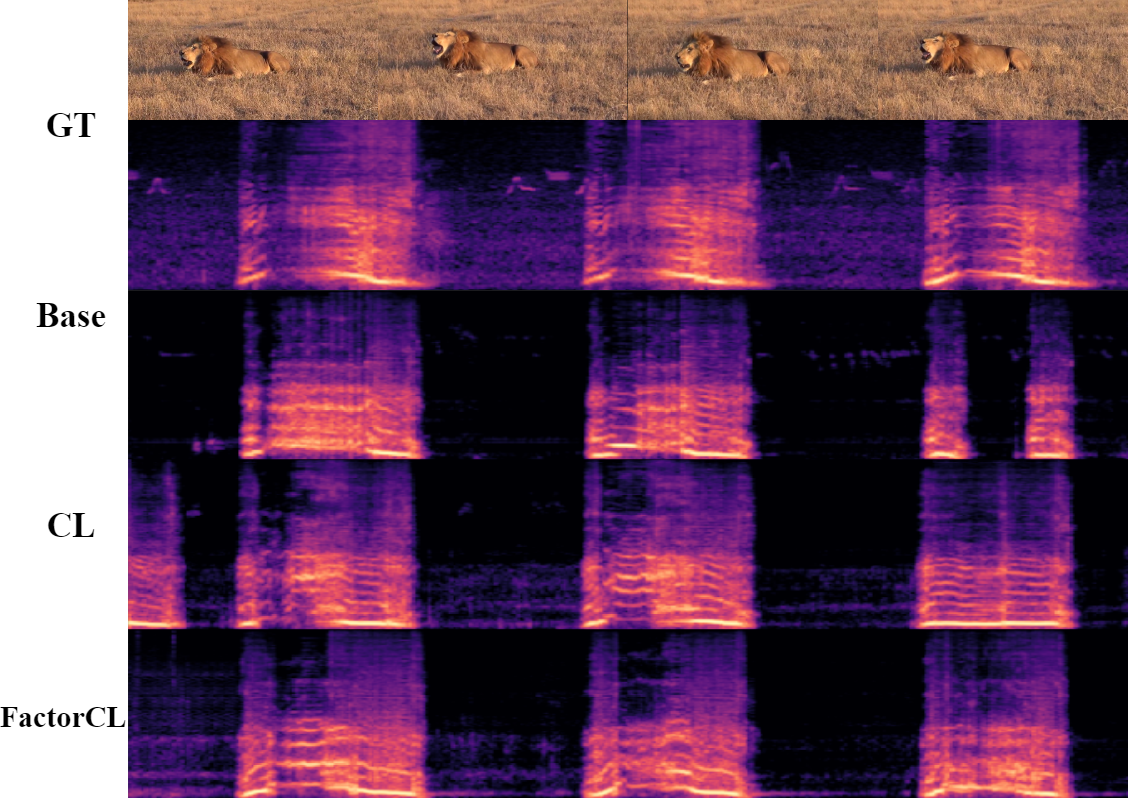}
    \caption{Mel spectrogram example for contrastive learning in VGGsound test set.}
    \label{mel}
\end{figure}

\begin{figure}[!htb]
    \centering    \includegraphics[width=0.35\textwidth]{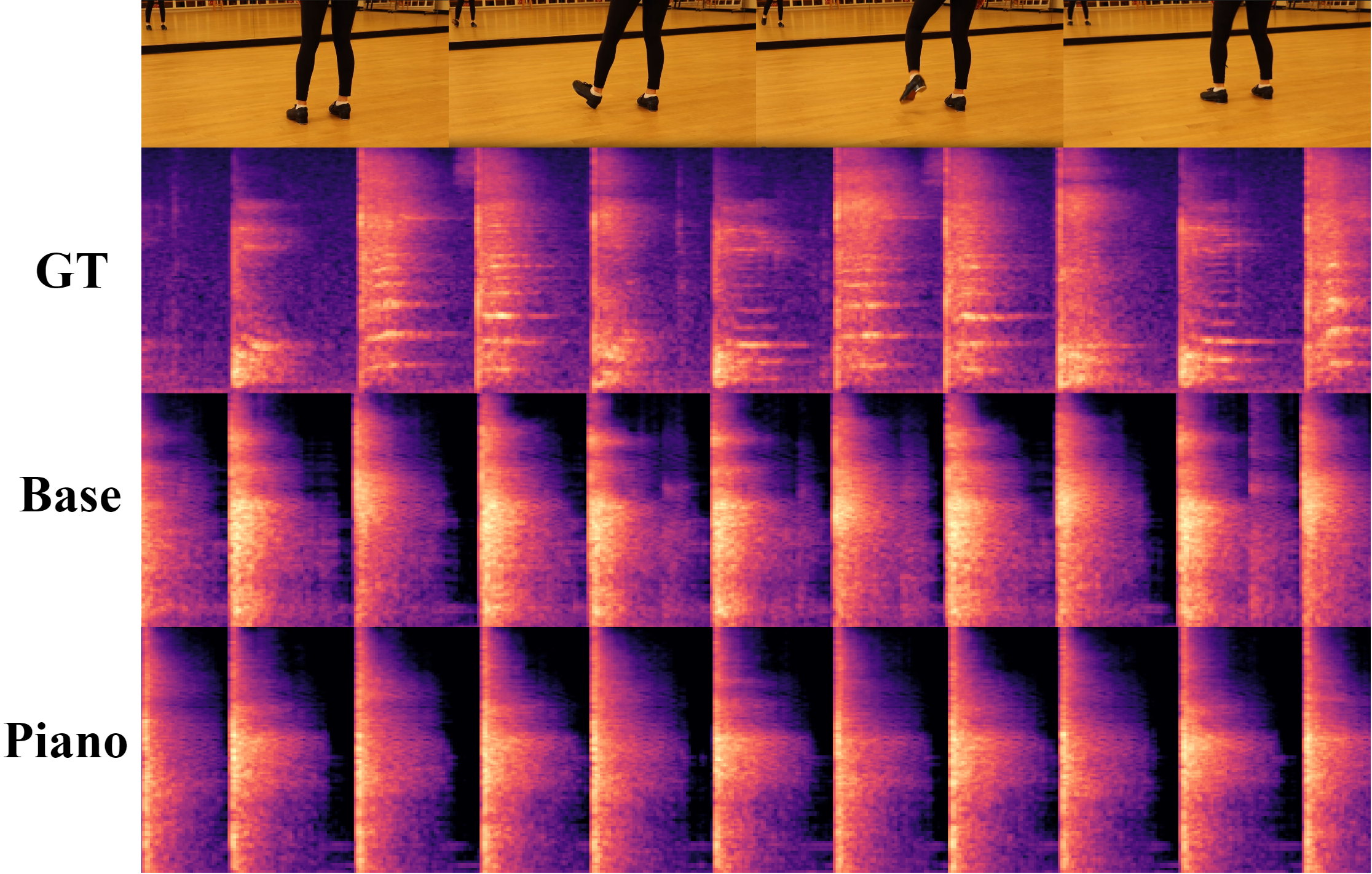}
    \caption{Mel spectrogram example for the piano model in VGGSound test set. The high similarity of Mel spectrograms demonstrates that our piano model maintains general V2A capabilities.}
    \label{piano}
\end{figure}

\begin{table*}[!htb]
\centering
\resizebox{0.9\textwidth}{!}{%
\begin{tabular}{lcccc}
\toprule
Method & SI-SDR$\uparrow$ & Melody Similarity (MOS)$\uparrow$ & Smoothness and Appeal (MOS)$\uparrow$ & MIDI Precision/Recall/Acc/F1 \\
\midrule
Audeo~\cite{Su2020AudeoAG} & 1.98 dB & 2.63 ± 0.10 & 3.25 ± 0.09 & \multirow{3}{*}{0.65/0.70/0.51/0.60} \\
Ours-Piano2h w/o guid. & -2.26 dB & 1.68 ± 0.11 & 3.29 ± 0.11 & ~ \\
Ours-Piano2h & \textbf{3.35 dB} & \textbf{3.44 ± 0.12} & \textbf{3.54 ± 0.11} & ~ \\
\bottomrule
\end{tabular}
}
\caption{Objective and subjective evaluations on PianoYT-2h. Frame-level MIDI precision, recall, accuracy, and F1 scores are computed following Audeo.}
\label{tab:piano2h}
\end{table*}

\begin{table*}[!h]
\centering
\resizebox{0.9\textwidth}{!}{%
\begin{tabular}{lcccc}
\toprule
Method & SI-SDR$\uparrow$ & Melody Similarity (MOS)$\uparrow$ & Smoothness and Appeal (MOS)$\uparrow$ & MIDI Precision/Recall/Acc/F1 \\
\midrule
Ours-Piano10h & 2.22 dB & 3.18 ± 0.12 & 2.96 ± 0.11 & \multirow{3}{*}{0.41/0.47/0.31/0.37} \\
Ours-Piano10h w. velocity & 3.06 dB & 3.14 ± 0.09 & 3.28 ± 0.11 & ~ \\
Ours-Piano10h w. velocity w. DPO & \textbf{3.21 dB} & \textbf{3.52 ± 0.08} & \textbf{3.32 ± 0.10} & ~ \\
\bottomrule
\end{tabular}
}
\caption{Objective and subjective evaluations on Piano-10h.}
\label{tab:piano10h}
\end{table*}

\section{Experiments}
\subsubsection{Datasets} For general audio generation, we use about 1.2M text-audio pairs, including AudioCaps~\cite{kim2019audiocaps}, WavCaps~\cite{mei2024wavcaps}, TangoPromptBank~\cite{ghosal2023tango}, MusicCaps~\cite{agostinelli2023musiclm}, and AF-AudioSet~\cite{gemmeke2017audio,kong2024improving}. We only use the VGGSound~\cite{chen2020vggsound} dataset as video-related data. The data sets are detailed in Table~\ref{tab:datasets}. For video-to-piano sound generation, We use datasets PianoYT-2h, as utilized by Audeo~\cite{Su2020AudeoAG}, and the front view of video and precise MIDI information of our Piano-10h.

\subsubsection{Implementation details} 
To improve training stability, multi-modal sound generation capability is progressively incorporated throughout the training process using a multi-stage training approach. The training is carried out with a batch size of 128 and a total of 330k steps on 8 Nvidia A800 GPUs. We use the Adam optimizer with a learning rate of 3e-5. We also clip the gradient norm to 0.2 for training stability. During inference, we use a sway sampling strategy~\cite{chen2024f5} with NFE $= 64$, and a classifier-free guidance strength of 2.0.

We reproduce Audeo's Video2Roll module using the same ResNet architecture and achieve very similar accuracy and recall rates comparable to those reported by Audeo. The piano model is further trained for an additional 4k steps, which prove sufficient for achieving convergence.

\subsubsection{Evaluation Metrics} 

We employ several metrics to evaluate semantic alignment, temporal alignment, and audio quality on the VGGSound test set and the AudioCaps test set, including Inception Score (IS)~\cite{salimans2016improved}, CLIP score, Fréchet Distance (FD)~\cite{heusel2017gans}, Fréchet Audio Distance (FAD), AV-align (AV)~\cite{yariv2024diverse}, KL Divergence-softmax (KL-softmax)~\cite{SpecVQGAN_Iashin_2021}, and CLAP score~\cite{wu2023large}.

For piano music evaluation, we calculate the Scale-Invariant Signal-to-Distortion Ratio (SI-SDR) in the frequency domain to measure the similarity between the generated music and the ground truth. We also conduct subjective evaluations comparing our model's generated piano music with Audeo's synthesized outputs. These evaluations assess both the similarity to the ground truth and the smoothness and musical appeal independent of the ground truth. Multiple evaluators provide Mean Opinion Scores (MOS) for these assessments.

\section{Results}
\subsection{T2A/V2A Foundation Model}
We compare our base model after training stage 1 on the VGGSound test set with those of existing state-of-the-art models. Our method outperforms the previous best with FAD $0.79$ to $0.74 (+6.33\%)$, CLIP Score $16.12$ to $17.70 (+9.80\%)$. The results are presented in Table~\ref{tab:vggsound}.

For T2A generation, as shown in Table~\ref{tab:audiocaps}, we follow the evaluation protocol of GenAU~\cite{haji2024taming} and MMAudio~\cite{cheng2024taming} and evaluate our base model on the Audiocaps test set. Our model still achieves competitive performance on T2A tasks.

\subsection{Piano Sound Generation}

As shown in Table~\ref{tab:piano2h}, our piano model consistently outperforms Audeo across both objective (SI-SDR $1.98 dB$ to $3.35 dB$ $+69.19\%$) and subjective (MOS $2.94$ to $3.49$ $+18.71\%$) metrics on the PianoYT-2h dataset, even though employing the same underlying feature extraction module. Meanwhile, it still retains good performance on the general V2A task.


\subsection{Ablation Studies}
We conduct a series of ablation experiments to demonstrate the effectiveness of our proposed multi-stage training and multi-step guidance dataset.

\begin{enumerate}

\item As shown in Table~\ref{tab:vggsound}, comparing \textit{Ours-Base w/o text} and \textit{Ours-Base w. text} reveals that incorporating textual conditions significantly enhances multi-modal audio generation. All metrics demonstrate improvements, with an average increase of $12.10\%$.

\item When the model is extended to \textit{Ours-Piano2h w. text}, as shown in Table~\ref{tab:vggsound}, an Extra-DiT is introduced to support music generation capabilities. All metrics remain comparable to those of \textit{Ours-Base w. text} on VGGSound, indicating that this extension achieves enhanced piano sound generation without substantially compromising the original V2A performance. Mel-spectrogram samples are shown in Fig.~\ref{piano}.

\item \textit{Ours-CL w. text} and \textit{Ours-FactorCL w. text} in Table~\ref{tab:vggsound} further validate the effectiveness of contrastive learning in multi-modal audio generation. Compared with \textit{Ours-Base}, \textit{Ours-FactorCL w. text} achieves an average improvement of $3.69\%$, showcasing the potential of factorized contrastive learning to improve both audio generation quality and cross-modal relevance. Mel-spectrogram samples are shown in Fig.~\ref{mel}.

\item As presented in Table~\ref{tab:piano2h} \textit{Ours-Piano2h w/o guid.}, we remove MIDI guidance and train both the Roll Predictor and the multi-stream DiTs in an end-to-end manner. The results demonstrate that a multi-step generation strategy is crucial for enhancing sound quality.

\item \textit{Ours-Piano10h w. velocity} and \textit{Ours-Piano10h w. velocity w. DPO} in Table~\ref{tab:piano10h} demonstrate the effectiveness of velocity guidance and DPO in enhancing learning performance style and audio quality. Compared to \textit{Ours-Piano10h w. velocity}, SI-SDR increases from $2.22 \,\mathrm{dB}$ to $3.21 \,\mathrm{dB}$ ($44.59\%$), Melody Similarity (MOS) improves from $3.18$ to $3.52$ ($10.38\%$), and Smoothness and Appeal (MOS) rises from $2.96$ to $3.32$ ($12.16\%$). These results highlight how additional guidance information and preference learning effectively boost music generation performance.

\end{enumerate}

\section{Conclusion} 
In this work, we propose a three-stage strategy for a flow matching based multi-stream DiTs architecture for multi-modal controlled sound generation. In the first stage, we train the base T/V2A model. In the second stage, an Extra-DiT is applied for specific tasks, such as piano music generation. In the third stage, post-optimization through contrastive learning and preference learning is employed to improve the generated audio quality. We achieve competitive results in both general sound generation tasks and specific tasks by our method. Additionally, we introduce a high-quality multi-view piano performance video dataset, which includes multi-view videos, highly accurate MIDI information, and diverse piano performance styles from different pianists. Using this dataset, we can improve the generation quality by leveraging various piano performance information, such as pitch, duration, velocity, and style, step by step in a CoT-like way. Evaluation results highlight the advantages of the proposed multi-stage generative framework compared to the state-of-the-art models on a variety of datasets, with FAD $0.79$ to $0.74$ ($+6.33\%$), CLIP $16.12$ to $17.70$ ($+9.80\%$) on VGGSound, SI-SDR $1.98 dB$ to $3.35 dB$ ($+69.19\%$), MOS $2.94$ to $3.49$ ($+18.71\%$) on PianoYT-2h, and SI-SDR $2.22 dB$ to $3.21 dB$ ($+44.59\%$), MOS $3.07$ to $3.42$ ($+11.40\%$) on Piano-10h.


\bibliographystyle{named}
\bibliography{ijcai25}
\end{document}